# SCREAMm – Modified Code SCREAM to Simulate the Acceleration of a Pulsed Beam through the Superconducting Linac[*]

Yu. Eidelman[#], S. Nagaitsev, N. Solyak, FNAL, Batavia, IL 60510, U.S.A

## Abstract

The code SCREAM [1,2] – **S**uper**C**onducting **RE**lativistic particle **A**ccelerator si**M**ulation was significantly modified and improved. Some misprints in the formulae used have been fixed and a more realistic expression for the vector-sum introduced. The realistic model of Lorentz-force detuning (LFD) is developed and will be implemented to the code. A friendly GUI allows various parameters of the simulated problem to be changed easily and quickly. Effective control of various output data is provided. A change of various parameters during the simulation process is controlled by plotting the corresponding graphs "on the fly". A large collection of various graphs can be used to illustrate the results.

## KEY FEATURES OF THE CODE

Code SCREAMm is used for the simulation of the longitudinal beam dynamics in a superconducting linac. To optimize the acceleration of a beam passing through a string of superconducting cavities the control of the phases and amplitudes of the RF fields in the cavities is required. This is particularly true in a section of the linac, where the beam velocity changes continuously, while the cavity-$\beta$ changes (only in a few discrete steps in a case of low beta sections). The RF control system becomes more complicated in the typical case in which multiple cavities are driven by one klystron. The changing beam velocity also results in increasing beam loading along the cavity string. It demands additional adjustment of the cavity fields. In addition, superconducting cavities are very susceptible to deformations of their shape caused by electromagnetic pressure (Lorentz-force detuning) and microphonics. Other effects which have to been taken into account include incoming beam energy/phase jitter, particle distributions in the bunch over energies and phases. All these and some other effects were taken into account in the code to design the realistic RF control system.

The SCREAMm program is a tool to calculate the optimal phase and amplitude settings for the superconducting cavities in a linac. It implements fast RF control techniques such as vector-sum regulation feedback. The earlier version contained a possibility to simulate fast ferrite vector-modulators [3] as part of the control system for Fermilab proton driver [4]. Each run in SCREAMm calculates the acceleration of the beam in one beam RF pulse. Several runs can be collected in one "file". The beam pulse (or run) contains many bunches. The beam pulse consists of two parts: cavity filling and flat-top interval. Obviously, the beam is launched only once filling is completed. To limit computing time, the particles in the bunch are regrouped into macro-particles, which have the mass and charge proportional to the sum of the particles in them but are regarded as single particles during the simulation of the acceleration process in the linac. The exact number of single particles contained in each macro-particle is taken into account when the beam loading is calculated. The macro-particles are distributed in longitudinal phase-space so as to simulate the distribution of the particles in a real bunch. Although more granular, bunches in the SCREAMm occupy more or less the same footprint in phase-space as the real bunches. The bunch-charge, bunch-centroid and arrival time at injection are randomly varied bunch-to-bunch ("incoherent") and pulse-to-pulse ("coherent") to explore the longitudinal acceptance of the linac. The program also provides an option of simulating different particle species.

## HOW PROGRAM SIMULATES THE LINAC

A code starts with reading of the structure of the linac from the input file. Once that is done, program sequentially calculates the energy/time profile of each macro-particle along the linac. This requires the simulation of the effective accelerating voltage provided by each cavity to each macro-particle. This voltage is not only determined by the particular cavity design field, by also by the phase difference of the each particle with respect to the cavity RF phase. Furthermore, voltage attenuation and phase error ensue in the case of detuning of the cavity, i.e. a shift of the cavity resonance frequency from the (fixed) klystron frequency as a result of mechanical deformations related LFD and microphonics.

Important effect of the mismatch between cavity-$\beta$ design and current particle velocity is described by the transit time factor. This effect is especially important in long multi-cell cavities. It also includes the effect of the sinusoidal variation of the accelerating field in the cavity in time and space on the effective acceleration of the beam.

Then typically the beam phase in proton linac is offset by $\approx$-10º to obtain gradient (or phase-) focusing of the bunch. Gradient focusing consists of accelerating the beam ahead of the RF crest so that slower particles are accelerated more than faster particles. Due to injection jitter, the particle phase difference smears the particle around the

___________________________________________

*Work supported by Fermi Research Alliance, LLC under contract DE-AC02-07CH11359 with the U.S. Department of Energy

#eidelyur@fnal.gov.

particle phase advance setting. The phase-factor is also strongly affected by detuning and beam loading in the cavities. Because of that variations of the amplitude and phase of the cavity accelerating voltage appear, which are cumulated in the acceleration history of each particle.

Once the current beam phases are calculated, the induced field (or beam loading) is subtracted from the cavity field. The refill of the cavity is simulated assuming constant klystron power. The power forwarded from klystron to the cavities (and reflected from them) is adjusted (with feed-forward and feedback) so as to keep the cavity voltage and phase constant. The feed-forward settings take into account the expected beam loading and LFD. Since RF-modules typically consist of a few cavities, special techniques are needed to regulate the RF power and phase in the individual cavities, which may differ by their detuning and beam loading. At the klystron level the phase and amplitude of the RF signal are typically regulated using vector-sum regulation. It consists of summing the vectors determined by a phase (vector direction) and amplitude (vector length) measured for each cavity, with a goal to derive the klystron RF phase and amplitude settings that produce the desired sum-vector. A more consistent expression (as in the old version) for power settings is used in the actual code, in particular, transit time factor is contained in it explicitly.

Constant phase offset settings such as the synchronous phase and beam phase advance are programmed into the low level RF system that drives the klystron RF phases with respect to the master oscillator (taking into account waveguide length differences as well as the three-stub tuner for the cavity-to-cavity differences within a module). The program obbiously assumes that the synchronous phase was determined and all phases used are therefore determined with respect to this phase.

With a typical time-step $1\mu s$, a thousand bunches are simulated for a typical beam-pulse of a few ms duration. A bunch moving with speed of light travels 300 m during $1\mu s$. A bunch is therefore usually at the end of the linac before next one is launched. The code therefore describes the progression of a bunch through the entire linac within one time step. Obviously many macro-particles of which the bunch consists are also tracked independently through the linac. The cavity field calculations (as well as update of the feedback loop, etc.) are performed once per time step, hence the linac field configuration is essentially static during the passage of the bunch. The previous bunches affect the actual bunch only through the accumulated beam-loading voltage.

No wake-field effects, space-charge effects as well intra-beam scattering effects between macro-particles or bunches are considered.

The program SCREAMm operates in a MATLAB environment. Using the numerous advantages of this package, the program unfortunately inherits its shortages – MATLAB is an interpreted language. For this reason, the code uses some C-language routines in the tracking part of the program to speed up the calculation time.

## INPUT/OUTPUT FEATURES

The input into code consists of defining several MATLAB structure variables and was initially provided in the form of Excel spreadsheet. SCREAMm inherits this approach but a special GUI window is used to edit input data file (Figure 1).

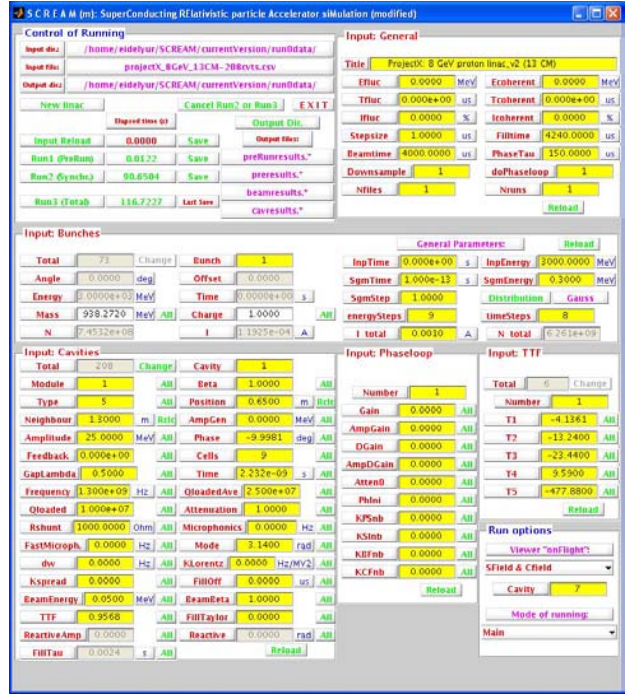

Figure 1. Main GUI window of code SCREAMm.

This window provides a quick and effective change of any general parameters as well as parameters describing each of the resonators, macro-particles and etc. Easy management of the calculations is provided in the code due to realized opportunity to observe the different output results "on the fly" (see, for example, Figure 2).

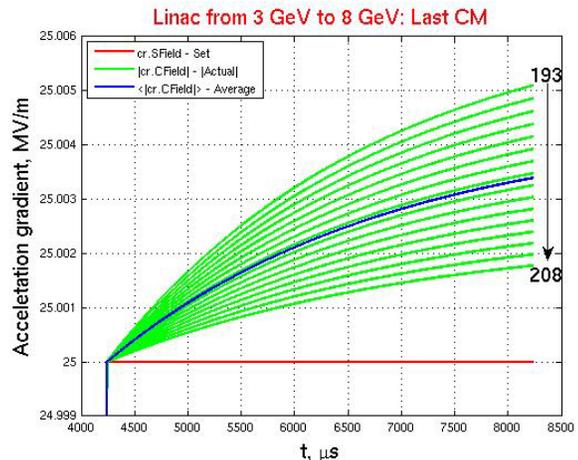

Figure 2. Accelerating gradient (green), vector-sum (blue)

and Set-point (red) vs time for last 16 cavities of the 8 GeV proton linac (ideal case: no errors, no LFD, no micriphonics).

Some other possibilities are realized due to GUI window: management of input/output files, "new linac creation", advance control of output data, the selection of different mode of the running and so on. A special MATLAB script was written as well to provide different output plottings.

## LORENTZ-FORCE DETUNING

Superconducting cavities are known to have a very narrow bandwidth (1 kHz or less) and therefore taking into account cavity Lorentz-force detuning is very important (figure 3).

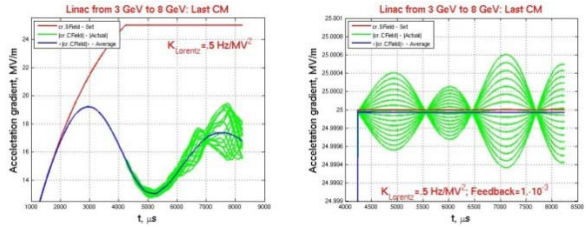

Figure 3. Accelerating gradient (green), vector-sum (blue) and Set-point (red) in the last 16 cavities (8 GeV proton linac), powered by one klystron in presence of LFD (old approach to calculate LFD) [6]. Left: in case of feedback OFF; right: feedback ON. Note that the scales are different. LFD Coefficient=-0.5 Hz/(MV/m)$^2$, Feedback gain is 100.

The code (in its early version) uses the following simplest expression to calculate value $\Delta\omega$ of LFD:

$$\Delta\omega = 2\pi \frac{\Delta t}{\tau} KV^2, \quad (1)$$

where $\Delta t$, $\tau$, $K$ and $V(t)$ are time step, some characteristic time (order of hundreds of microseconds), Lorentz coefficient and actual acceleration gradient in cavity. This approach does not take into account different "strength" $K_m$ of individual mechanical eigenmodes of the cavities with frequencies $\omega_m$ and quality factors $Q_m$. Therefore a more realistic model of LFD was developed. In this model the following equation for detuning $\Delta\omega$ was investigated:

$$\Delta\ddot{\omega} + \frac{\omega_m}{Q_m}\Delta\dot{\omega} + \omega_m^2 \Delta\omega = 2\pi\omega_m^2 K_m V^2(t) \quad (2)$$

Presence of quality factor leads to the splitting of the eigenmode frequency:

$$\omega_{1,2} = i\frac{\omega_m}{2Q_m} \pm \frac{\omega_m}{2Q_m}\sqrt{4Q_m^2 - 1} \equiv i\gamma_m \pm \overline{\omega}_m. \quad (3)$$

Equation (2) was solved for three time intervals: filling stage $(0 \le t \le \tau_{fill})$, flat-top period $(\tau_{fill} \le t \le T)$ and RF-off stage $(t > T = \tau_{fill} + \tau_{beam})$. During first stage the acceleration voltage increases linearly with time; when a beam passes through the cavity (second stage) the acceleration voltage keeps the constant value $V_0$ over time and after that decreases exponentially (last interval). Although the proposed model is quite simple, but expressions for the detuning frequency are somewhat cumbersome:

$$\Delta\omega_{fill}(t) = 2\pi K_m \left(\frac{V_0}{\tau_{fill}}\right)^2 \left\{ t^2 - \frac{2t}{\omega_m Q_m} - 2\frac{Q_m^2 - 1}{\omega_m^2 Q_m^2} + \right.$$
$$\left. 2\frac{e^{-\gamma_m t}}{\omega_m^2 Q_m^2}\left[(Q_m^2 - 1)\cos\overline{\omega}_m t + \frac{3Q_m^2 - 1}{\sqrt{4Q_m^2 - 1}}\sin\overline{\omega}_m t\right]\right\},$$

$$\Delta\omega_{flat-top}(t) = 2\pi K_m V_0^2 \left\{1 - e^{-\gamma_m(t-\tau_{fill})} \cdot \right.$$
$$\left[\cos\overline{\omega}_m(t-\tau_{fill}) + \frac{\sin\overline{\omega}_m(t-\tau_{fill})}{\sqrt{4Q_m^2-1}}\right]\right\} + e^{-\gamma_m(t-\tau_{fill})} \cdot$$
$$\left\{\frac{\Delta\dot{\omega}_\tau + (\gamma_m + \overline{\omega}_m \tan\overline{\omega}_m\tau_{fill})\Delta\omega_\tau}{\overline{\omega}_m}\sin\overline{\omega}_m(t-\tau_{fill}) + \right.$$
$$\left.\frac{\Delta\omega_\tau}{\cos\overline{\omega}_m\tau_{fill}}\cos\overline{\omega}_m t\right\},$$

$$\Delta\omega_{RF-off}(t) = \frac{2\pi\omega_m^2 K_m V_0^2 e^{-\gamma_{RF} t}}{\gamma_{RF}^2 - 2\gamma_{RF}\gamma_m + \omega_m^2}\left\{1 - e^{(\gamma_{RF}-\gamma_m)(t-T)} \cdot \right.$$
$$\left\{\left[\cos\overline{\omega}_m(t-T) - \frac{(\gamma_{RF}-\gamma_m)}{\overline{\omega}_m}\sin\overline{\omega}_m(t-T)\right]\right\} +$$
$$+ e^{-\gamma_m(t-T)}\left\{\frac{\Delta\dot{\omega}_T + (\gamma_m + \overline{\omega}_m \tan\overline{\omega}_m T)\Delta\omega_T}{\overline{\omega}_m} \cdot \right.$$
$$\left.\sin\overline{\omega}_m(t-T) + \frac{\Delta\omega_T}{\cos\overline{\omega}_m T}\cos\overline{\omega}_m t\right\};$$

$\Delta\omega_{\tau,T}$, $\Delta\dot{\omega}_{\tau,T}$ are detuning and its derivatives reached at the appropriate time ($\tau_{fill}$ or $T$, correspondingly) and $\gamma_{RF} = \omega_{RF}/2Q_{RF}$ characterises the decay of the RF field in the cavity after switching off the klystron power.

The expressions obtained for LFD were used to compare the predictions of the proposed model with experimental data [5] for a TESLA type cavity. The example of such comparison is shown in Figure 4. A different number of eigenmodes was taken into account. Results of this study show that more than ten eigenmodes must be taken into account to match the experimental data. The spectrum and parameters of these modes were determined based on the analysis of numerous experimental data [7]. The approach developed for LFD calculation will be implemented to the code SCREAMm.

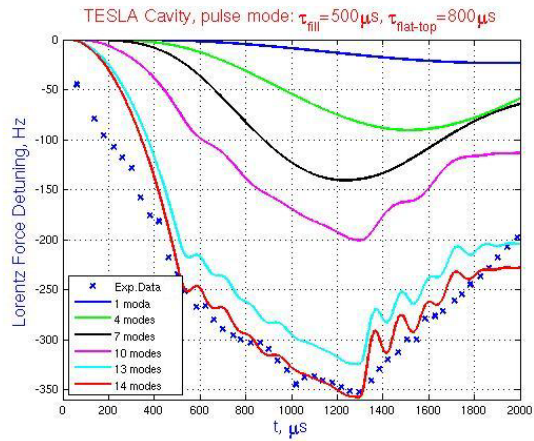

Figure 4. Lorentz-force detuning: comparison experimental data and simulation results.

## CONCLUSIONS

The existing code SCREAM was modified. New features of the program have significantly improved its prospects for solving various problems of simulation of a superconducting linac. The new code SCREAMm is used to optimize the parameters of the 8 GeV linac as part of Project X developed in Fermilab.